\documentclass[pra,twocolumn,a4paper,showpacs]{revtex4}

\usepackage[english]{babel}
\usepackage{graphicx}
\usepackage{amsmath,amssymb,amsfonts}
\usepackage{hyperref}
\usepackage{bm}

\begin{document}

\title{Tailoring of motional states in double-well potentials by time-dependent processes}

\date{\today}

\author{Kari~H\"{a}rk\"{o}nen}
\email{kari.harkonen@utu.fi}
\author{Ollijuhani~K\"{a}rki}
\author{Kalle-Antti~Suominen}
\affiliation{Department of Physics, University of Turku, FI-20014 Turku, Finland}

\begin{abstract}
We show that the vibrational state tailoring method developed for molecular systems can be applied for cold atoms in optical lattices. The original method is based on a three-level model interacting with two strong laser pulses in a counterintuitive sequence [M.~Rodriguez \textsl{et al.}, Phys. Rev. A \textbf{62}, 053413 (2000)]. Here we outline the conditions for achieving similar dynamics with single time-dependent potential surfaces. It is shown that guided switching between diabatic and adiabatic evolution has an essential role in this system. We also show that efficient and precise tailoring of motional states in optical lattices can be achieved, for instance, simply by superimposing two lattices and moving them with respect to each other.
\end{abstract}


\maketitle


\section{Introduction}

Complete control of atomic center-of-mass motion is one of the main principles of physics with cold atoms~\cite{CohenTannoudji1992,Metcalf1999} and ions~\cite{Ghosh1995,Leibfried2003}. Cooling of atomic samples allows their trapping with far-off resonant optical traps~\cite{Grimm2000} or with spatially inhomogeneous magnetic fields~\cite{Pethick2002}. This has made it possible, for example, to explore the quantum statistics of bosonic and fermionic atoms~\cite{Pethick2002,Greiner2003}, to develop ultrahigh precision spectroscopy and atomic clocks~\cite{Bize2005}, and to produce cold molecules by photoassociation~\cite{Weiner1999}. One specific example is an optical lattice, where the internal state structure of atoms and the polarization states of light combine to produce periodic and controllable potentials for atoms~\cite{Jessen1996}. By sufficient cooling the atoms can be localized into the lattice sites. The situation is reminiscent of solid state systems, and interesting effects such as formation of a Mott insulator~\cite{Greiner2002,Morsch2006} or the surface of a Fermi sea~\cite{Koehl2005,Bloch2005} have been observed. This has motivated ideas of building quantum simulators for solid state phenomena, and even quantum computers~\cite{Lloyd1996,Cirac2003}. For such purposes one needs a wide-ranging and versatile toolbox for the quantum control of atomic motion. 

In optical lattices atoms can either move in the Bloch bands of the periodic structure, or become localized at lattice sites (not forgetting the interesting intermediate region where the motion or ``hopping" takes place by quantum tunnelling between the lattice sites). Here we consider the case of rather deep localization, where the eigenstates of the atomic center-of-mass states are quite discrete in energy and become the vibrational states of a single lattice site. Our aim is to develop a tool for moving atoms efficiently and selectively to these vibrational states if they are initially in the vibrational ground state. Especially, we want to consider the case where the state change is associated with a move from one lattice site to another.

We approach the problem with ideas originally proposed for the quantum control of vibrational states in dimer molecules~\cite{Garraway1998, Rodriguez2000, Garraway2003}. Note that for atoms the vibrational states are center-of-mass states, whereas in molecules they correspond to the internal motion, i.e., relative motion of the nuclei.  An efficient method to move molecules between vibrational states is STIRAP (stimulated raman adiabatic passage)~\cite{Vitanov2001}, which is based on the existence of a noninteracting eigenstate in a three-level system and a counterintuitive pulse sequence. The advantage of the method is its high efficiency (population transfer is nearly complete), robustness in relation to the parameters, and the elimination of the intermediate state. For molecules as well as for atoms in a lattice the intermediate state would usually be an electronically excited one, which normally decays rapidly and incoherently. 

The standard STIRAP approach, however, is strongly dependent on overlap integrals between the vibrational states, which in molecules means favorable Franck-Condon factors. This limits the achievable ``stretching" of the molecules as the process needs to adhere to the Franck-Condon principle. In addition, any implementation for optical lattices would require additional pulsed laser fields. An interesting and viable alternative in molecules is offered by very strong pulses, which actually affect the vibrational state structure~\cite{Giusti-Suzor1995}. As demonstrated in Ref.~\cite{Garraway1998}, this situation can be modelled with LIP's, i.e., light-induced potentials, which are the time-dependent potential surfaces associated with the instantaneous eigenstates of the combined system of three electronic states and two pulses. In fact, the process is described by adiabatic evolution along a single LIP, which efficiently takes the molecule from one electronic state to another. As the equilibrium positions associated with each electronic state may be quite different, the process leads to an apparent breakdown of the Franck-Condon principle. Later it was shown that with a proper setting of parameters one can decide whether the vibrational state is changed during the process as well~\cite{Rodriguez2000}.

Here we show that the same method is applicable for atoms in optical lattices or in double-well structures. First we demonstrate that under suitable conditions we can indeed reduce the system of three states and two pulses into a single time-dependent potential. As explained in Sec.~\ref{sec:derivation}, this is not so straightforward as one could assume. The fact that the process is not completely adiabatic (as noted in Ref.~\cite{Rodriguez2000}) plays a role here, as well as in tuning the parameters to achieve the population of a specific vibrational state. In Sec.~\ref{sec:results} we provide some numerical results that demonstrate the practicality of the approach, and in Sec.~\ref{sec:2D} we address specifically the two-dimensional case (the original molecular studies were performed in one dimension). A special case of two optical lattices that can be moved with respect to each other is treated in Sec.~\ref{sec:lattice}. Finally, we conclude our presentation in Sec.~\ref{sec:conclusions} with a discussion on applications of the method.


\section{\label{sec:derivation}Three-state and single-state descriptions}

In case of molecular dimers, the system is originally described in terms of a three-component state vector consisting of electronic state wave functions~\cite{Rodriguez2000}. Next we present a highly simplified picture of a dimer system. The advantage is that the description of the system can be generalized beyond the dimer concept to any suitable three-state case, and then reduced to a single-state model. Both cases, three state and single state, are in principle applicable to suitable optical lattice systems, and furthermore, the latter model can be also used to describe situations which are genuinely single-state cases, i.e., there is no underlying three-state structure present.

\begin{figure}[tb!]
\includegraphics[width=70mm]{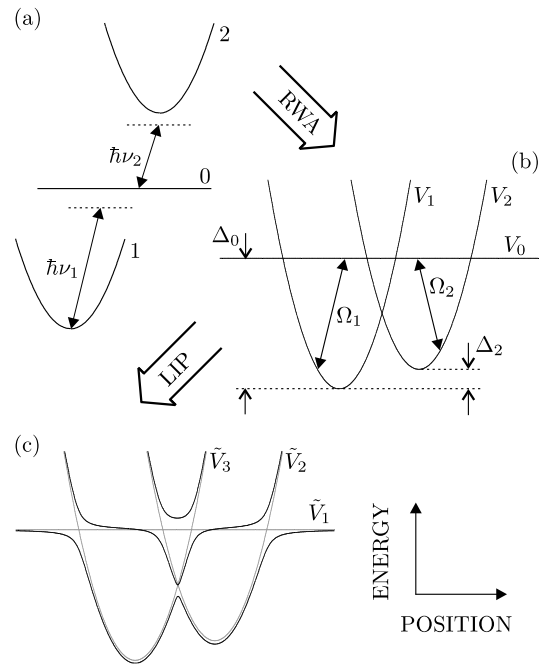}
\caption{\label{fig:schematicPotentials}(a) The initial system consists of three individual potential surfaces coupled to each other via two pulsed lasers with frequencies $\nu_{i}$. (b) In the rotating wave approximation (RWA) the system is characterized by Rabi frequencies $\Omega_{i}$ and detunings $\Delta_{i}$. (c) Finally, the light-induced potentials (LIP's) are derived from the RWA potential matrix, Eq. \eqref{eq:dimerpot}, as its time and position dependent eigenvalues, shown here by dark lines.}
\end{figure}

Let us denote the state vector components $\psi_{i}(\mathbf{x},t),i=1,0,2$ (following Ref.~\cite{Rodriguez2000}). Consequently, the time evolution of such a state vector $\Psi(\mathbf{x},t)$ is given by the scaled time-dependent Schr\"{o}dinger equation
\begin{eqnarray}   
   i \frac{\partial}{\partial t} \Psi(\mathbf{x},t)=-\nabla^{2} \Psi(\mathbf{x},t)
   +\mathcal{V}  (\mathbf{x},t)\Psi(\mathbf{x},t),\label{eq:ourschrodinger}
\end{eqnarray}   
where $\mathbf{x}$ and $t$ are the scaled position and time. Because of the used scaling, every quantity is expressed with a dimensionless number, and especially $\hbar=1$ and $m=1/2$. The associated electronic potentials and state couplings form the matrix  
\begin{eqnarray}\label{eq:dimerpot}
   \mathcal{V}(\mathbf{x},t)=
   \left(
   \begin{array}{c c c}
      V_{1}(\mathbf{x}) & \Omega_{1}(t) & 0 \\ 
      \Omega_{1}(t) & V_{0}(\mathbf{x})  & \Omega_{2}(t)\\
      0 & \Omega_{2}(t) & V_{2}(\mathbf{x})\\
   \end{array}
   \right),
\end{eqnarray}
where 
\begin{eqnarray}
   V_{i}(\mathbf{x})&=&\textstyle \frac{1}{4} \displaystyle \omega_{i}^{2} (\mathbf{x}-\mathbf{x}_{i})^{2}+\Delta_{i},\quad i=1,0,2,\\
   \Omega_{j}(t)&=&\Omega_{j}^{0}\exp[-(t-t_{j})^{2}/T_{j}^{2}],\quad j=1,2. 
\end{eqnarray}
In other words, we assume either harmonic or flat potentials and Gaussian pulses. Also, we have above applied the rotating wave approximation (RWA), which allows us to shift potentials appropriately. Thus the potentials $V_{i}(\mathbf{x})$ are parameterized by trapping frequencies $\omega_{i}$ and detunings $\Delta_{i}$, and the coupling terms $\Omega_{j}(t)$ include maximum Rabi frequencies $\Omega_{j}^{0}$ and time scales for the pulses $T_{j}$. The role of parameters and the effect of RWA is demonstrated in Fig.~\ref{fig:schematicPotentials}.

We assume that $\Delta_{1}=0$ and $\omega_{0}=0$. This assumption is justified because the effect of $\Delta_{1}$ is merely a universal shift in the energy levels and the shape of the potential $V_{0}$ does not have any significant effect on the results~\cite{Rodriguez2000}. We are interested in adiabatic time evolution with the exception that time evolution may include sudden rapid diabatic behavior. The potential matrix~\eqref{eq:dimerpot} can always be diagonalized and we obtain three real eigenvalues as a result. Note that these eigenvalues depend both on time and position; they can be considered as new potentials in the eigenstate basis. We label these light-induced potentials (LIP's) with $\tilde{V}_{i}(\mathbf{x},t)$. The eigenstate basis turns out to be equivalent to the adiabatic basis and the original three-state basis is called the diabatic basis. The numbering of the LIP's goes as 1, 2, and 3, with increasing energy.

In our model the couplings $\Omega_{i}$ in the diabatic basis are initially (for $t\ll t_{1}, t_{2}$) and finally (for $t\gg t_{1}, t_{2}$) practically but not quite zero. Then the diabatic and adiabatic potentials are very much alike, except that any degeneracy is a level crossing for the first case, and an avoided crossing for the second case, see Fig.~\ref{fig:schematicPotentials}(c). If we are initially on the diabatic state $1$, in the molecular model this means that adiabatic evolution will keep us in the appropriate electronic eigenstate, which at the final point turns out to have evolved into the diabatic state 2. For a system with three states this works very well, but if we merely take the eigenstate potential $\tilde{V}_{1}(\mathbf{x},t)$ and solve the appropriate single-state Schr\"{o}dinger equation, we find that initially our system corresponds to a double-well potential [combination of $V_{1}(\mathbf{x})$ and $V_{2}(\mathbf{x})$; here $\mathbf{x}_{1}$ and $\mathbf{x}_{2}$ are assumed to be clearly different]. As the right-hand well begins to move down in energy initially due to the strengthening coupling pulse $\Omega_{2}$, the separating energy barrier is too weak to hold the initial state localized to the potential well $V_{1}(\mathbf{x})$, see again Fig.~\ref{fig:schematicPotentials}(c). In the three-state model this problem does not arise, because initially the system is held by the diabatic potential $V_{1}(\mathbf{x})$.

\begin{figure}[tb!]
\includegraphics[width=75mm]{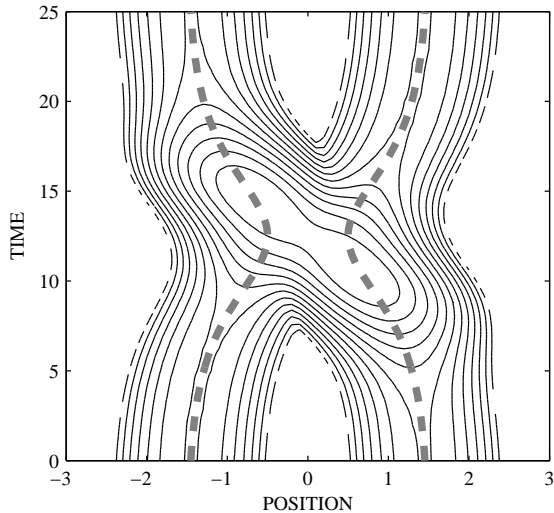}
\caption{\label{fig:shiftingOfTheWells}The time evolution of the light-induced potential surface $\tilde{V}_{1}(\mathbf{x},t)$. For clarity, the energy values have been cut from above. Here we have $\Delta_{0}>0$ and thus the flat potential surface is located above the centers of the two harmonic ones, as in Fig.~\ref{fig:schematicPotentials}(c). Due to the counterintuitive pulse order ($t_{1}>t_{2}$), the right-hand-side well drops down first, but the initially large position separation keeps the system in the diabatic state 1 long enough. The grey dashed lines mark $\mathbf{x}_{1}(t)$ and $\mathbf{x}_{2}(t)$.}
\end{figure}

In other words, the apparent adiabatic evolution along a single LIP is possible in the model only because in reality the evolution is diabatic initially and finally, and adiabaticity sets in only in the middle of the process. In order to work with truly single-state models, we need to modify the molecular approach. An obvious solution is to make $\mathbf{x}_{1}$ and $\mathbf{x}_{2}$ time dependent as well, by setting their initial and final values sufficiently large, so that the double-well potential barrier is strong when the adiabatic description would otherwise fail, see Fig.~\ref{fig:shiftingOfTheWells}.

\section{\label{sec:results}Numerical results for the single-state model}


\subsection{Description of numerical methods}

We consider the single-state wave function as a wave packet of plane waves and evolve it in a discretized space numerically with the methods of wave packet dynamics~\cite{Garraway1995}. We solve the time-dependent Schr\"{o}dinger equation using the split-operator and fast Fourier transformation (FFT) methods. In addition, for a better understanding of dynamics we follow Ref.~\cite{Rodriguez2000} and solve also the time-independent Schr\"{o}dinger equation for fixed moments of time using the Numerov algorithm~\cite{Pang1997}. We simplify our notation by defining 
\begin{eqnarray}
	\psi(\mathbf{x},t) & \equiv &\psi_{1}(\mathbf{x},t),\\
	\tilde{V}(\mathbf{x},t) & \equiv & 
		\begin{cases}
			\tilde{V}_{1}(\mathbf{x},t), & \textrm{if $\Delta_{2}>0$},\\
			\tilde{V}_{2}(\mathbf{x},t), & \textrm{if $\Delta_{2}<0$}.
		\end{cases}
\end{eqnarray}
Obviously, it would be seemingly practical to set the three-state model completely aside instead of using it as a tool for obtaining the single-state LIP, but with this approach we can use the knowledge obtained with previous studies of the molecular system.

To model the time dependence of $\mathbf{x}_{1}$ and $\mathbf{x}_{2}$, we redefine our diabatic potentials $V_{1}$ and $V_{2}$ as
\begin{eqnarray}
  V_{j}(\mathbf{x},t)&=&\textstyle\frac{1}{4}\omega_{j}^{2}[\mathbf{x}-\mathbf{x}_{j}(t)]^{2}
  +\Delta_{j},\\ 
  \mathbf{x}_{j}(t)&=&
	\begin{cases}
		\alpha_{j}+\beta_{j}\,\text{sech}[\gamma_{j}(t-t_{s})], & \textrm{(1D)} \\
		\big(\,0,\,\alpha_{j}+\beta_{j}\,\text{sech}[\gamma_{j}(t-t_{s})]\,\big), & \textrm{(2D)}
	\end{cases}
\label{eq:shifting}
\end{eqnarray}  
where $t_{s}$ is the moment when the individual potential wells are closest to each other (Fig.~\ref{fig:shiftingOfTheWells}). For simplicity we use in our calculations equal maximum Rabi frequencies $\Omega_{1}^{0}=\Omega_{2}^{0}\equiv\Omega$ and equal time scales of the pulses $T_{1}=T_{2}\equiv T$. Pulses are applied in a counterintuitive order, that is $t_{1}>t_{2}$. We choose the time $t_{s}$ so that $t_{s}=\frac{1}{2}(t_{1}+t_{2})$. Since the individual potential wells are initially apart from each other, we can consider them effectively as two separate harmonic potentials, and as a consequence, we select our initial state $\psi(\mathbf{x},t_{0})$ to be case specifically the ground or excited state of the harmonic
potential $V_{1}$. 


\subsection{\label{subsec:1Dcase}One-dimensional case}

In our scaled units time $t$ evolves from values $0$ to $25$ and position ranges from $-3$ to $3$. We choose the moments for the pulse maxima as $t_{1}=15$ and $t_{2}=10$. The pulse time scales $T$ are set to $T=5$. The trapping frequencies of the harmonic potentials $V_{1}$ and $V_{2}$ are $\omega_{1}=\omega_{2}=20$. As for the shifting parameters, we set $\alpha_{1}=-\alpha_{2}=-1.5$, $\beta_{1}=-\beta_{2}=1.0$, and $\gamma_{1}=\gamma_{2}=0.30$ [cf. the trajectories of $\mathbf{x}_{j}(t)$ in Fig.~\ref{fig:shiftingOfTheWells}]. As stated earlier, we have set $\omega_{0}=0$ and $\Delta_{1}=0$.

\begin{figure}[tb!]
\includegraphics[width=80mm]{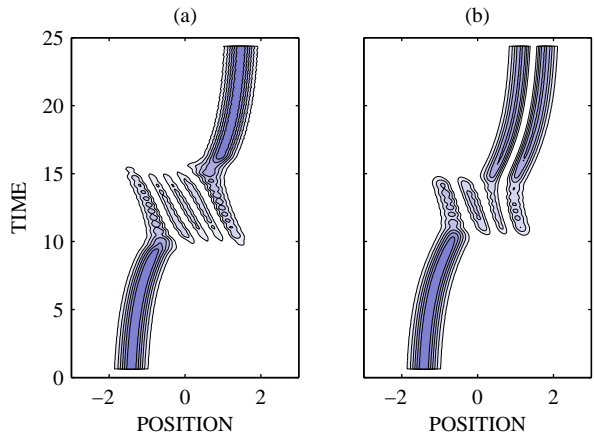} 
\caption{\label{fig:evolution1D}(Color online) (a) Time evolution for a process that drives population from the state corresponding the left ground state into that of the right one (cf. corresponding potential surface in Fig.~\ref{fig:shiftingOfTheWells}). Here $\Delta_{0}=850$, $\Delta_{2}=0$, and $\Omega= 274$. (b) An alternative process where the excited vibrational states is reached. Here $\Delta_{0}=850$, $\Delta_{2}=-20$, and $\Omega= 162$.}
\end{figure}

The above settings leave as control parameters $\Delta_{0}$, $\Delta_{2}$, and $\Omega$. If we set $\Delta_0=850$, $\Delta_{2}=0$, and $\Omega = 274$, we obtain the basic case of a transfer from the ground vibrational state of potential $V_{1}$ to the ground vibrational state of potential $V_{2}$ in the three-state model. The evolution is plotted in Fig.~\ref{fig:evolution1D}(a). The change from one ground state to another one is almost complete. 

As an alternative selection we can take $\Delta_{0}=850$, $\Delta_{2}=-20$, and $\Omega = 162$. Figure~\ref{fig:evolution1D}(b) shows that the final state becomes the first excited state of potential $V_{2}$, as expected for the analogous three-state case.

\begin{figure*}[bt!]
\includegraphics[width=170mm]{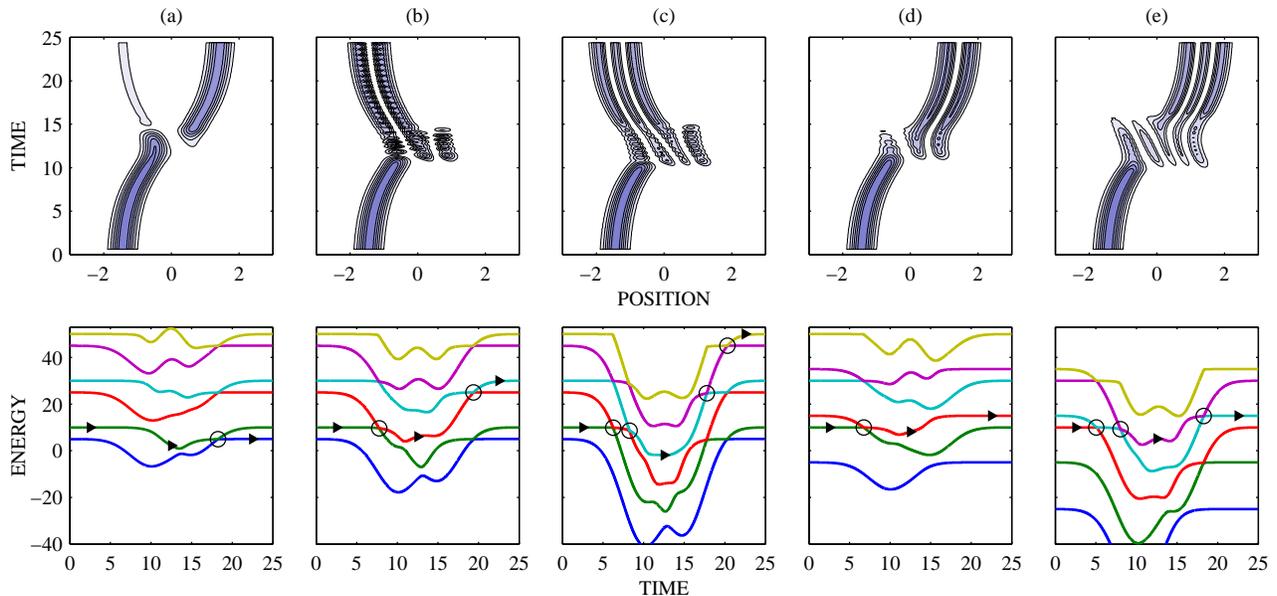} 
\caption{\label{fig:delta2nonzero}(Color online) Examples of time evolution (upper panels) and  corresponding energy spectra of six lowest LIP eigenstates (lower panels) for different parameter values. We have chosen $\Delta_{0}=850$ in every frame while $\Delta_{2}$ and $\Omega$ are varied, such that (a) $\Delta_{2} = -5$, $\Omega = 100$, (b) $\Delta_{2} = -5$, $\Omega = 140$, (c) $\Delta_{2} = -5$, $\Omega = 200$, (d) $\Delta_{2} = -15$, $\Omega = 100$, and (e) $\Delta_{2} = -35$, $\Omega = 175$. Other parameters are as mentioned in Sec.~\ref{subsec:1Dcase}. In the energy spectra, the populated state is marked with arrows and the encountered diabatic jumps with circles.}
\end{figure*}

By further changing parameter values it is possible to achieve transfer from the ground state of $V_{1}$ to various excited states of $V_{2}$, as shown in Fig.~\ref{fig:delta2nonzero} (upper panels). As in the three-state case, it is also possible to return completely to the state $V_{1}$, but with a different vibrational state. The initial state can also be a vibrationally excited one for $V_{1}$. One can also make the transfer incomplete, which produces well-to-well superposition states. Note also, by comparing Fig.~\ref{fig:evolution1D}(a) with Fig.~\ref{fig:delta2nonzero}(a), and Fig.~\ref{fig:evolution1D}(b) with Fig.~\ref{fig:delta2nonzero}(d), that the number of peaks at the midpoint of evolution can be different even if the initial and final states are the same. We return to this issue in Sec.~\ref{sec:vibrational}.

\subsection{\label{subsec:roleOfParameters}The role of parameters} 

The number of parameters in the full three-state model is very large: maximum Rabi frequencies~$\Omega_{j}^{0}$, detunings~$\Delta_{i}$, trapping frequencies $\omega_{i}$, time scales of pulses~$T_{j}$, and pulse delay $\Delta t=t_{1}-t_{2}$. Furthermore, one can consider asymmetric pulse shapes, chirped pulses or some different potential geometry. Consequently, the parameter space is too large for a complete mapping, and we limit our discussion to the effect of $\Delta_{0}$, $\Delta_2$, and $\Omega$ within the previous parameter settings. With these three parameters, we are already able to target arbitrary vibrational states of both $V_{1}$ and $V_{2}$.

The sign of $\Delta_{0}$ is essential. Choosing $\Delta_{0}<0$, i.e., locating $V_{0}(\mathbf{x})$ below the other two curves, will result in the very robust APLIP process~\cite{Garraway1998}, in which the wave packet is smoothly transferred from one ground state to another, but no other tailoring is possible. The opposite selection leads to much richer dynamics and excitations become possible as well~\cite{Rodriguez2000}, but we pay for it with the dynamics becoming far more sensitive to the parameters. In short, as long as $|\Delta_{0}|$ is large enough, its sign is a sufficient switch between the two cases. From an alternative point of view, changing the sign of $\Delta_{0}$ is also equivalent to swapping the pulses $\Omega_{1} \leftrightarrow \Omega_{2}$.

The role of $\Omega$ is intuitively clear: the more powerful are the couplings, the more are the adiabatic potentials modulated. On the other hand, $\Omega$ and $|\Delta_{0}|$ counteract each other, such that the larger $|\Delta_{0}|$, the larger values of $\Omega$ are needed to modify the potentials by the required amount. Of course, they do not exactly cancel each other, since lower values of $|\Delta_{0}|$ tend to allow the wells to broaden, too. Nevertheless, the cancellation is efficient from the point of view of regions close to the minima of the wells, where the dynamics is also concentrated. Therefore, the counterplay between the detuning $\Delta_{0}$ and the coupling strength $\Omega$ explains, e.g., the plotted stripe structure of Fig.~3 in Ref.~\cite{Garraway2003}. We have verified the existence of these stripes also in our single-state model. 

\begin{figure}[tb!]
\includegraphics[width=85mm]{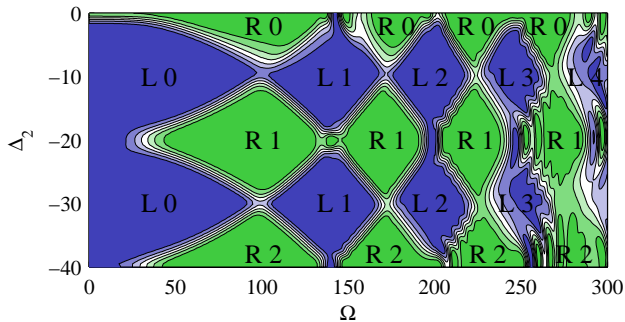} 
\caption{\label{fig:deltavsomega}(Color online) The final states as a function of $\Omega$ and $\Delta_{2}$. In the labels \textsl{R} and \textsl{L} signify whether the wave packet resides at the end in the right-hand or left-hand well, respectively, and the number indicates the vibrational quantum number of the final state.}
\end{figure}

The detuning $\Delta_{2}$ between potentials $V_{1}$ and $V_{2}$ has an obvious function. If the potentials $V_{j}$ are to be considered one by one, they each have their own (and in this case the same) frequency $\omega$. Because of our scaling, the corresponding energy difference between the harmonic eigenstates is $\Delta E = \omega$ in the scaled units. Starting with $\Delta_{2}=0$, we can find such combinations of parameters that will lead to the transfer of the population from one ground state to another as in Fig.~\ref{fig:evolution1D}(a). The basic idea is that lowering $V_{2}$ by $\Delta_{2}=-\Delta E$ will drive the population now into the first excited state with approximately the same parameters [cf.~Fig.~\ref{fig:evolution1D}(b)]. This generalizes to higher excited states as well: $\Delta_{2} = -n\times\Delta E$, $n\in \mathbb{N}$, will lead to the $n$th excited state of potential 2. Interestingly, by setting $\Delta_{2} = -(n+1/2)\times\Delta E$ and keeping $\Omega\leq 100$ we always get the initial state, i.e., lowest vibrational state of potential 1 back, but by increasing $\Omega$ we can reach the higher vibrational states of potential 1. So the tuning process is clearly sensitive to the value of $\Omega$, as shown in Fig.~\ref{fig:deltavsomega}. 

In the mapping of Fig.~\ref{fig:deltavsomega} we can see that the final outcome forms regular patterns. Especially, clear systematicity arises. On the one hand, higher vibrational states of $V_{1}$ are gained by increasing $\Omega$ while $\Delta_{2}$ acts only in a periodic way. On the other hand, the higher vibrational states of $V_{2}$ are achieved by decreasing $\Delta_{2}$ while $\Omega$ has no significant role. 


\subsection{\label{sec:vibrational}Interpretation in vibrational basis}

\begin{figure*}[tb!]
\includegraphics[width=170mm]{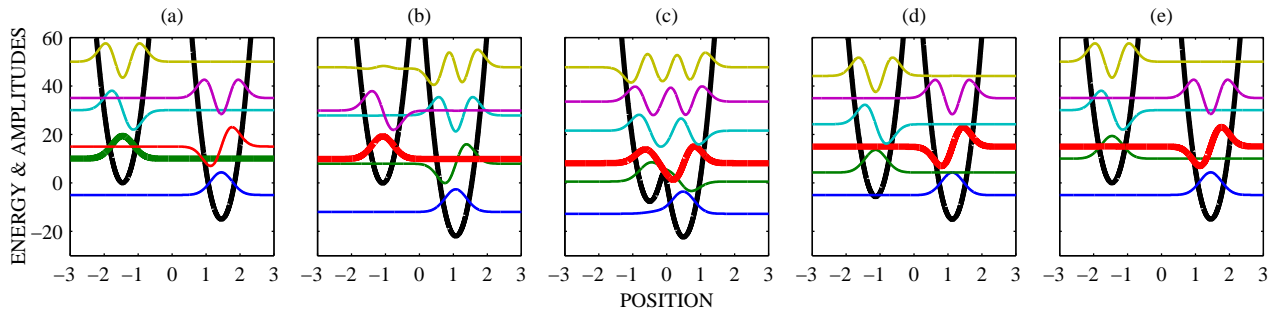} 
\caption{\label{fig:vibrationalBasisVibLIP}(Color online) The evolution of the double-well potential surface and the corresponding LIP eigenstates. This series corresponds the choice of aprameters in Fig. \ref{fig:delta2nonzero}(d). Each frame represents a time slice, such that (a) $t=0$, (b) $t=7.5$, (c) $t=12.5$, (d) $t=18$, and (e) $t=25$. For clarity, the amplitudes have been rescaled and the populated state is plotted with a thicker line. A rapid diabatic jump has occured at $t \approx 7$.}
\end{figure*}

In order to understand the role of the parameter choices and the dynamics in general, we have followed Ref.~\cite{Rodriguez2000} and analyzed the situation by 
solving the instantaneous vibrational states of the relevant LIP (we call them LIP eigenstates). If the two wells are sufficiently separated, the lowest few LIP eigenstates are very much similar to those of two single harmonic potentials $V_{1}$ and $V_{2}$, except for the narrow degenerated cases, when the energies of two LIP eigenstates coincide. As the energy of a LIP eigenstate gets comparable to the height of the separating potential barrier, it starts to have non-negligible component in both wells, as seen in Fig.~\ref{fig:vibrationalBasisVibLIP}. 

During the pulses the double-well potential gets modulated in such a manner, that the potential barrier becomes lower or even disappears. Accordingly, the set of corresponding LIP eigenstates changes as well. As the height of the barrier gradually decreases, the minima of the two wells move with respect to each other in energy. With large enough values of $\Omega$, the changes are considerable, if compared to the energy separations $\Delta E$. Consequently, in terms of the LIP eigenstates, the states approximately corresponding to the single-well eigenstates appear to cross each other. This phenomenon can be easily seen in the energy spectra (cf. Fig.~\ref{fig:delta2nonzero}). Of course, there is always a small energy gap between the energies, i.e., the crossings are in fact avoided.

The sharp crossings are an indication of diabatic evolution. Let us define crossing time scale 
\begin{equation}
\tau_{\text{cross}}^{n} \equiv \min\big\{t'>0 \big|\,\, |\langle\phi_{n}(t) | \phi_{n+1} (t + t') \rangle |^{2} \sim 1 \big\}, 
\end{equation}
where $\phi_{n}(t)$ is the $n$th LIP vibrational eigenstate. With the given choice of parameters, the transformations from one eigenstate to another happen in time scale of $\tau_{\text{cross}}^{n}\sim 10^{-2}$, which is less than the vibrational time scale of the potential $\tau_{\text{vib}}\sim 10^{-1}$. Accordingly, the wave packet cannot succeed in following these changes in the vibrational basis, and therefore a diabatic jump between the LIP eigenstates occurs. 

Between the sharp diabatic crossings, the system follows a particular LIP eigenstate adiabatically. Adiabatic following occurs when the energy gap between the two approaching LIP eigenstates is large. Consequently, in case there are many possible energy level crossings, the time scale of potential modulations $\tau_{\text{pot}}$ can be used as a switch to determine which crossings are passed diabatically and which adiabatically. By making a crossing into a mixture of diabatic and adiabatic, one can reach superpositions, although the precise result is likely to be very sensitive on parameter values.    

It is worthwhile to emphasize, that the seemingly rapid and radical changes in the appearence of the wave packet typically around $t\approx 10$ and $t\approx 15$ do not have anything to do with diabatic crossings. In Fig.~\ref{fig:vibrationalBasisVibLIP} we have plotted the LIP eigenstates corresponding the choice of parameters in Fig.~\ref{fig:delta2nonzero}(d). By comparing these figures, it is obvious, that the diabatic jump occurs already at $t\approx 7$, when nothing seems to happen in the time evolution. Interestingly, the subsequent changes in the course of evolution are in fact due to the evolution of the corresponding LIP eigenstate itself. Therefore, all the visible transformations are actually a manifestation of adiabatic following. 

Returning to the discussion of Fig.~\ref{fig:deltavsomega} in the preceding section, we are now ready to give explanation to the pattern. Let us denote the number of sharp LIP energy level crossings by $(n,m)$, where $n$ and $m$, respectively, count the crossings before and after the pulse maxima. For example, the situation in Fig.~\ref{fig:delta2nonzero}(a) corresponds to the notation $(0,1)$. 

\begin{figure}[tb!]
\includegraphics[width=80mm]{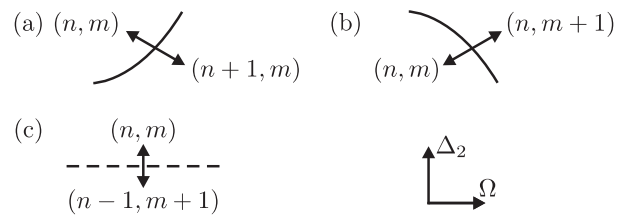} 
\caption{\label{fig:crossingsAtTheEdge}The three possible types of changes in the numbers of sharp LIP energy level crossings before and after pulse maxima $(n,m)$, respectively, at the edges of the zones in Fig.~\ref{fig:deltavsomega}. (a) Positive and (b) negative slopes affect different numbers. (c) An additional rule associated with the resonance energies $\Delta_{2}\equiv 0 \pmod{ \Delta E}$ (dashed line; not a visible edge).}
\end{figure}

The edges of the zones arise for two reasons. First, those with a positive slope occur because a new LIP energy level crossing is being formed prior to the pulses, i.e., $(n,m) \leftrightarrow (n+1,m)$ at the edge. Second, those with a negative slope are associated with a creation of a crossing after the pulses, i.e., $(n,m) \leftrightarrow (n,m+1)$ at the edge. Yet another rule for crossings can be formulated, but it does not correspond to any visible edge in the mapping. The third rule is, that when passing the resonance detuning $\Delta_{2}\equiv 0 \pmod{ \Delta E}$ from above, one crossing travels via infinity from left to right, i.e., $(n,m) \leftrightarrow (n-1,m+1)$. Illustration of the rules is given in Fig.~\ref{fig:crossingsAtTheEdge} (cf. spectra in Fig.~\ref{fig:delta2nonzero} and the mapping in Fig.~\ref{fig:deltavsomega}). 

With the three rules, we can see that $|n-m| \le 1$. If $n=m$, then the $n$th vibrational state of potential $V_{1}$ is achieved. If $n=m\pm 1$, then a state in  $V_{2}$ is reached, such that each increment of $-\Delta E = -20$ in $\Delta_{2}$ introduces one step higher excitations. Advancing towards greater values of $\Omega$ brings along more diabatic jumps. Naturally, the stability suffers as the complexity of the process increases, which can be seen as a degeneration of the pattern with large values of $\Omega$. 


\subsection{\label{sec:2D}Two-dimensional case}

So far we have studied only the one-dimensional case. In two dimensions, we set the potentials symmetric for simplicity and apply a time-dependent shift to the potentials $V_{1}$ and $V_{2}$. The shifting is the same as previously in the one-dimensional case, but it is conducted only in one direction, as defined in Eq.~\eqref{eq:shifting}. 

\begin{figure}[tb!]
\includegraphics[width=80mm]{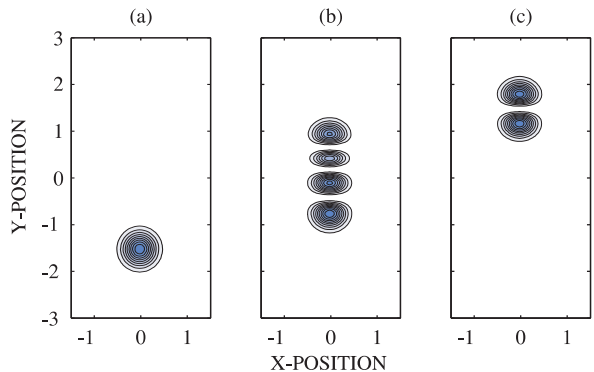} 
\caption{\label{fig:evolution2D}(Color online) The double-well time evolution in two dimensions. The frames show the stage of evolution at (a) $t=0$, (b) $t=10$, and (c) $t=20$. The parameters are the same as for the corresponding one-dimensional case in Fig.~\ref{fig:evolution1D}(b).}
\end{figure}

There is no coupling between the Cartesian coordinates $x$ and $y$ in the potentials $V_{1}$ and $V_{2}$. Accordingly, Eq.~\eqref{eq:ourschrodinger} is separable in $x$ and $y$. Thus the two-dimensional system is in practice split into one-dimensional slices which are labelled by $x=$ constant. The only difference between two $x=$ constant slices is a shift in energy. Therefore, we can expect the system to behave in a similar way as a one-dimensional system. Our numerical results confirm this, see Fig.~\ref{fig:evolution2D}. 

We have used the same parameter values in our calculations as in the corresponding one-dimensional case. It should be emphasized, though, that the two-dimensional system is much more sensitive to specific parameter values than the one-dimensional system. Another difference is the fact that we can apply the time-dependent shift to the potentials $V_{1}$ and $V_{2}$ only in one direction~\cite{torus}. Thus the possible excited state outcomes of the processes have excited vibrational states in that particular direction only. For optical lattice configurations that opens interesting possibilities. For example, one can shift the potentials in $y$ direction in the first process and then take the final state of that as the initial state of another process where the roles of $y$ and $x$ are reversed.


\section{\label{sec:lattice}Application to a modulated lattice}


\subsection{The basic idea}

The preceding model discussed so far originates from a three-state model and therefore inherits a somewhat unintuitive parametrization by pulse strengths and detunings. The actual geometry of the double-well potential is given as a solution of a third degree polynomial resulting from an eigenvalue calculation of the matrix given in Eq.~\eqref{eq:dimerpot}. The solution cannot be simplified into any illustrative and compact form in terms of the control parameters, which is also why it is not given explicitly in the text, and mainly qualitative characterizations could be given in Sec.~\ref{subsec:roleOfParameters}. Nevertheless, the previously derived one-state model is adequate for presenting the general behavior of wave packet dynamics in time-dependent double wells. Therefore, the results do not depend on the actual formulation of the underlying potential. 

We now move on to seek alternative realizations for the double-well potential. The main focus is to make the control of the potential as self-explanatory as possible. Another point of interest is to introduce a method which could be also experimentally accessible. 

\begin{figure}[tb!]
\includegraphics[width=80mm]{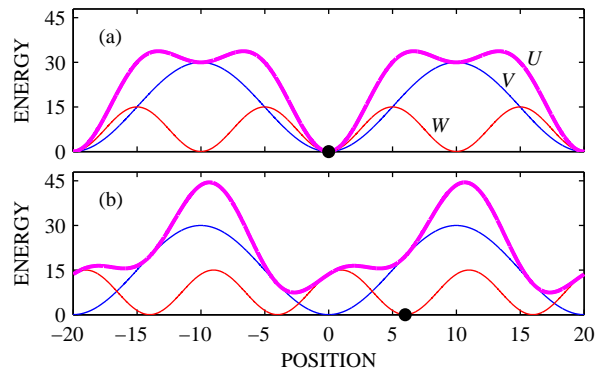} 
\caption{\label{fig:latticeModulationPotentials}(Color online) The sinusoidal potentials $V(x)$ and $W(x,t)$ used in the lattice modulation model. Superposition of these two results in a lattice of modulated double-well potentials $U(x,t)$, as $W$ moves with respect to $V$. The strengths of the lattice potentials are chosen such that $V_{0} = 2 W_{0} = 30$. The dot marks the adjustable lattice offset parameter $x_{0}$. The frames represent the situation with (a) $x_{0}=0$ and (b) $x_{0}=6$.}
\end{figure}

The previous dynamics included typically two sets of diabatic crossings: one before the pulse maxima and another one after them. We now propose a possible scheme which will include only one set of energy level crossings. With the interest in optical lattices, we formulate the problem in terms of sinusoidal potentials. The modulation of the potential is done by another sinusoidal potential (but with a doubled frequency), which can be moved with respect to the other one (cf. Fig.~\ref{fig:latticeModulationPotentials}). Consequently, each site of the potential lattice will now become a time-dependent double-well potential.

Let the wavelengths of the potentials be $\lambda_{i}$ in scaled units. Then the respective potentials are
\begin{subequations}
\begin{eqnarray}
V(x) &=& V_{0} \sin^{2} ( 2 \pi x / \lambda_{V}), \\
W(x,t) &=& W_{0} \sin^{2} \{ 2 \pi [x - x_{0}(t)] / \lambda_{W} \}, \\
U(x,t) &=& V(x) + W(x,t),
\end{eqnarray}
\end{subequations}
where $\lambda_{V} = 2 \lambda_{W}$ and $x_{0}(t)$ describes the offset of $W$ with respect to $V$. The initial position is chosen such that the minima of both $V$ and $W$ coincide, i.e., $x_{0}(0) = 0$. Furthermore, approximately $V_{0} \approx 2 W_{0}$ is needed. 

The scaling is chosen as previously with an added assumption that $\lambda_{V} = 2 \lambda_{W} = 40$. Accordingly, the recoil energies $E_\textrm{recoil}^{i} = 4 \pi^{2} \hbar^{2} /(2 m \lambda_{i}^{2})$ of the lattice fields become $E_{\textrm{recoil}}^{V} = \pi^{2}/400$ and $E_{\textrm{recoil}}^{W} = \pi^{2}/100$ in the scaled units.

For simplicity, let us consider only one lattice site located initially at $x=0$. At first, the wave packet is at rest at the bottom of the well. Thereafter, the $W$ potential sweeps one-half of its wavelength to the right. Meanwhile, the wave packet gets driven uphill by the moving potential barrier. However, since the barrier effectively becomes lower after having passed the $x=0$ point, the wave packet will drop down to a vibrational state of a newly created well around $x=0$. 

\begin{figure}[tb!]
\includegraphics[width=80mm]{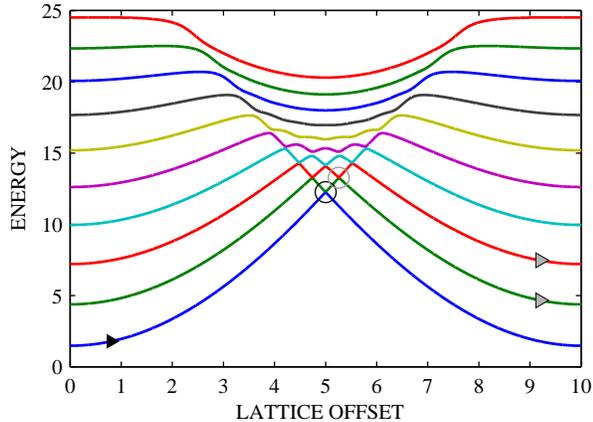} 
\caption{\label{fig:energySpectrumLatticeModulation}(Color online) The energy spectrum during lattice modulation as a function of the lattice offset parameter $x_{0}$. Here $V_{0}=30 \approx 1220\,\, E_{\textrm{recoil}}^{V}$ and $W_{0}=15 \approx 152\,\,E_{\textrm{recoil}}^{W}$ are used. The initial state is marked with an arrow. Using a sweep rate as in Fig.~\ref{fig:evolutionLatticeModulation}, a superposition of the first two excited states (grey arrows) is achieved by passing the first crossing diabatically (black circle) and the second partially diabatically (grey circle).}
\end{figure}

The point is that from the perspective of the initial ground state there will be only one set of energy level crossings during the period, when the wave packet is driven upwards in a local minimum, while a new global minimum is deepened on the other side of the barrier. The energy gaps in these crossings increase with the corresponding vibrational state number (cf. Fig.~\ref{fig:energySpectrumLatticeModulation}). Therefore, the sweep rate $\tau_{\text{pot}}$ will determine which crossings are passed diabatically, and which are followed adiabatically. 

\subsection{Numerical results}

Sinusoidal potentials are relatively flat and round if compared to the LIP's described in Sec.~\ref{sec:derivation}. Therefore, the corresponding double-well structure holds a set of eigenstates, which do not represent eigenstates of individual wells very well, but are spread over both wells instead. Consequently, in case of the higher vibrational states the energy level crossings are not as sharp as previously. Accordingly, the population is after the sweep likely to be spread over a distribution of vibrational states, as the population leaks gradually at each energy level crossing. Alternatively, a high control over the sweep rate $\tau_{\text{pot}}(t)$ is needed in order to slow down at the instant when adiabatic following during a crossing is wanted, and accelerate again thereafter in order to cross the other possible gaps diabatically.  

\begin{figure}[tb!]
\includegraphics[width=80mm]{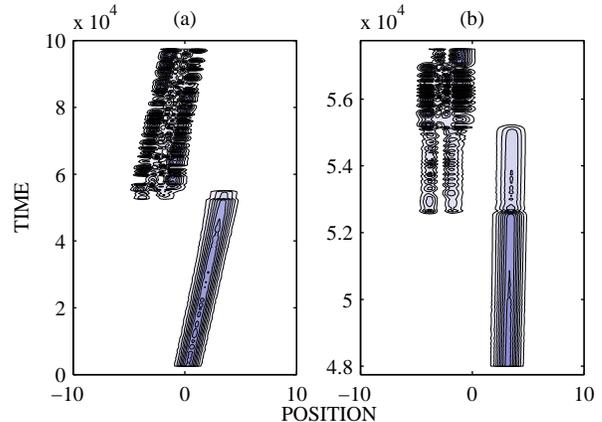} 
\caption{\label{fig:evolutionLatticeModulation}(Color online) Time evolution during the lattice modulation. The second plot is a high-resolution view of the transition moment, and illustrates the speed of the process compared to the smoother behavior in the harmonic wells. The lattice parameters are chosen as in Fig.~\ref{fig:energySpectrumLatticeModulation} and a constant sweep rate is applied.}
\end{figure}

Figure~\ref{fig:evolutionLatticeModulation} shows an example of the lattice evolution corresponding to Fig.~\ref{fig:energySpectrumLatticeModulation}. The lattice offset is evolved linearly, i.e., $x_{0}(t) = t / \tau_{\text{pot}}$, where the constant sweep rate is $\tau_{\text{pot}} = 10^{4}$. A superposition of the  first two excited vibrational states is populated accordingly. The above-mentioned leaks of population to other vibrational states result in small oscillations of the final wave packet. The process of change is rather fast as expected, too. Nevertheless, this example shows clearly that the ideas developed for the three-state system with two pulses can be carried over to the optical lattices.


\section{\label{sec:conclusions}Discussion}

We have investigated a transition from a counterintuitively coupled three-state system into a single-state system described by a time-dependent double-well potential. Thereafter, we have shown how the populations of the vibrational states in each well can be tailored by modulations of this double-well potential. 

The focus of our study was in investigation of necessary diabatic crossings blended among  otherwise adiabatic following. The system holds basically three different time scales: (i)~the intrinsic vibrational time scale $\tau_{\text{vib}}$ induced by potential geometry, (ii)~the general time scale of the modulation of the potential surface $\tau_{\text{pot}}$, and (iii)~a set of time scales associated with the crossings of the LIP eigenstates $n\in \mathbb{N}$ with their upper neighbors $\{ \tau_{\text{cross}}^{n} \}$. First, $\tau_{\text{vib}} \ll \tau_{\text{pot}}$ is needed in order to have adiabatic following in general. Second, the tailoring of the state populations is done by adjusting the set $\{\tau_{\text{cross}}^{n}\}$ with respect to $\tau_{\text{vib}}$ such that certain energy gaps are crossed diabatically while the others are avoided adiabatically. The values of  $\tau_{\text{cross}}^{n}$ are ultimately determined by the way the potential geometry is modified, but $\tau_{\text{pot}}$ scales the overall time scale of the process. Therefore, the adjustment of crossing time scales can be made by tuning either the potential parameters or the general time scale $\tau_{\text{pot}}$. 

In this general model survey we have used scaled units rather than taking a specific physical system with fixed parameters. This was done intentionally in order to focus on describing the model itself. In practice, we expect that the results can be applied in various systems. A recent example is an optical lattice setup where one generates a periodic two-dimensional lattice of double-well structures, with controllable barrier heights and relative well depths~\cite{Sebby2006}.

Another interesting situation arises when we consider our model from the viewpoint of a moving atom in an atom chip waveguide~\cite{Folman2002}. The position in our description can be considered as the transverse direction of the waveguide, and the time evolution corresponds to the change in the potential seen by the atom during its motion in the longitudinal direction. Our model can then describe a controlled transfer process between two waveguides, with the additional possibility of exciting tranverse modes.

Our future investigations will consider the role of interactions in this process. The simplest system to study is a zero-temperature Bose-Einstein condensate. The interactions cause the LIP eigenstate structure to depend on the momentary local density and this back-action needs to be taken into account. Also, in two dimensions it will remove the separability of the coordinates.


\begin{acknowledgments}
The authors acknowledge the financial support by the Academy of Finland (projects 206108 and 105740) and by the Vilho, Yrj\"{o} and Kalle V\"{a}is\"{a}l\"{a} Foundation (KH).
\end{acknowledgments}



\begin{thebibliography}{<99>}
\bibitem{torus} If we consider, instead, our one-dimensional position coordinate as a transverse cut in the radial direction of two torus-shaped potentials with slightly differently located minima, we can create transversal excitations.
\end{thebibliography}
\end{document}